\documentstyle[aps,12pt]{revtex}

\begin{document}
\author{E.N. Glass\thanks{%
Permanent address: Physics Department, University of Windsor, Ontario N9B
3P4, Canada} and J.P. Krisch}
\address{Department of Physics, University of Michigan, Ann Arbor, Michgan 48109}
\date{December 21, 1998}
\title{Two-Fluid Atmosphere for Relativistic Stars}
\maketitle

\begin{abstract}
\\\\We have extended the Vaidya radiating metric to include both a radiation
fluid and a string fluid. This paper expands our brief introduction to
extensions of the Schwarzschild vacuum which appeared in 1998 Phys. Rev. 
{\bf D 57} R5945. Assuming diffusive transport for the string fluid, we find
new analytic solutions of Einstein's field equations.\newpage\ 
\end{abstract}

\section{INTRODUCTION}

Strings have become a very important ingredient in many physical theories.
They may have been present in the early universe and played a role in the
seeding of density inhomogeneities \cite{mit&tur}. The idea of strings is
fundamental to superstring theories \cite{sen}. String fluids have been
suggested as dark matter candidates. The lensing properties of cosmic
strings and string systems have been discussed \cite{gott} and the apparent
relationship between counting string states and the entropy of the
Schwarzschild horizon \cite{larsen},\cite{mald},\cite{thooft} suggests an
association of strings with black holes. Recently Glass and Krisch \cite
{ed&jean} have pointed out that allowing the Schwarzschild mass parameter to
be a function of radial position creates an atmosphere with a string fluid
stress-energy around a static, spherically symmetric, object. If the mass is
also a function of retarded time a Vaidya radiation fluid is present in
addition to the string fluid. Since the metric has one arbitrary function, $%
m(u,r)$, invariantly defined by the sectional curvature of the $u,r$
two-surfaces, a given mass distribution determines the stress-energy.

The string fluid stress-energy is a macroscopic, statistically averaged,
description of a microscopic distribution of Planck length string bits. The
string fluid lies on a two-dimensional timelike world-sheet ($u,r$) and its
stress-energy is parametrized by radial derivatives of the mass function.
The radial stress, $p_r<0$, is the expected string tension. Since the string
fluid is constrained to stay in the ($u,r$) plane, the Planck scale string
bits have only radial motions. Interactions of the string bits along the
radial directions are modeled by the macroscopic string tension $p_r$ and
create, on average, a long string stretching radially away from the core or
horizon of the mass distribution. The Glass-Krisch atmosphere allows
transverse stresses. These stresses, if present, can be modeled by the
presence of a pressureless dust fluid. The transverse pressures arise from
the different velocities of the dust and string fluids. The most general
atmosphere has three components: a Vaidya radiation fluid, a string fluid
with radial tension, and a dust fluid. This can be interpreted as a
two-fluid atmosphere with a transverse pressure. We also discuss other, less
general, possibilities.

This atmosphere can model a variety of physical situations at different
distance scales. It can describe the atmosphere around a black hole with a
distance scale of multiples of Schwarzschild radii. It could also describe a
globular cluster with a dark matter component and a distance scale of the
order of parsecs.

The contracted Bianchi identities are satisfied for arbitrary $m(u,r)$ and
so the mode of mass propagation is a modeling choice from the many possible
propagation equations of classical and quantum physics. Each choice allows
the generation of new analytic solutions to the field equations. In this
work we develop new solutions by considering diffusive transport of the
string fluid elements. With this choice we find that the string fluid
diffuses inward as the Vaidya photons carry energy outward. The overall
effect is to slow down the time scale of the Vaidya energy loss.

In the next section we present our extension of Vaidya's metric \cite{vaidya}
as a further extension of the Schwarzschild vacuum solution. The arbitrary
mass function, which appears in the metric, is identified as the sectional
curvature of nested two-spheres. The string fluid and an associated static
two-fluid model are described in the third and fourth sections. We are able
to provide a two-fluid kinematic interpretation of the transverse stresses
in this model. In the fifth part of the paper diffusive transport is
developed and analytic solutions for the energy density are found. In the
sixth section analytic mass solutions are presented and interpreted. The
last two sections contain the analysis of the horizon structure of some of
our solutions and a discussion of the general results.

In this work Greek indices range over ($0$,$1$,$2$,$3$) = ($u,r,\vartheta
,\varphi $). Our sign conventions are $2A_{\nu ;[\alpha \beta ]}=A_\mu R_{\
\ \nu \alpha \beta }^\mu ,$ and $R_{\mu \nu }=R_{\ \mu \nu \alpha }^\alpha .$
$\ $Overdots abbreviate $\partial /\partial u$, and primes abbreviate $%
\partial /\partial r$. Overhead carets denote unit vectors. We use units
where $G=c=1$. Einstein's field equations are $G_{\mu \nu }=-8\pi T_{\mu \nu
},$ and the metric signature is (+,-,-,-).

\section{EXTENDING THE SCHWARZSCHILD VACUUM}

The spacetime metric covering the region exterior to a spherical star is
given by 
\begin{equation}
ds^2=Adu^2+2dudr-r^2(d\vartheta ^2+\text{sin}^2\vartheta d\varphi ^2)
\label{met1}
\end{equation}
where $A=1-2m(u,r)/r$. Initially $m(u,r)=m_0$ provides the vacuum
Schwarzschild solution in the region $r>2m_0$. At later times that region
admits a two-fluid description of radial strings and outward flowing
short-wavelength photons (sometimes called a ''null fluid''). Metric (\ref
{met1}) is spherically symmetric and given in retarded time coordinate $u$.
With the use of a Newman-Penrose null tetrad the Einstein tensor is computed
from (\ref{met1}) and given by 
\begin{eqnarray}
G_{\mu \nu } &=&-2\Phi _{11}(l_\mu n_\nu +n_\mu l_\nu +m_\mu \bar{m}_\nu +%
\bar{m}_\mu m_\nu )  \label{ein1} \\
&&-2\Phi _{22}l_\mu l_\nu -6\Lambda g_{\mu \nu }.  \nonumber
\end{eqnarray}
Here the null tetrad components of the Ricci tensor are 
\begin{mathletters}
\begin{eqnarray}
\Phi _{11} &=&(2m^{\prime }-rm^{\prime \prime })/(4r^2),  \label{ein1a} \\
\Phi _{22} &=&-\dot{m}/r^2,  \label{ein1b} \\
\Lambda &=&R/24=(rm^{\prime \prime }+2m^{\prime })/(12r^2).  \label{ein1c}
\end{eqnarray}
The only non-zero component of the Weyl tensor is 
\end{mathletters}
\begin{equation}
\Psi _2=-m/r^3+(4m^{\prime }-rm^{\prime \prime })/(6r^2).  \label{psi2}
\end{equation}
The metric is Petrov type {\bf D} with $l_\mu $ and $n_\mu $ principal null
geodesic vectors 
\begin{mathletters}
\begin{eqnarray}
l_\mu dx^\mu &=&du,  \label{nteta} \\
n_\mu dx^\mu &=&(A/2)du+dr,  \label{ntetb} \\
m_\mu dx^\mu &=&-(r/\surd 2)(d\vartheta +i\ \text{sin}\vartheta d\varphi ),
\label{ntetc}
\end{eqnarray}
where 
\end{mathletters}
\begin{mathletters}
\begin{eqnarray}
l_{\mu ;\nu } &=&(A^{^{\prime }}/2)l_\mu l_\nu -(1/r)(m_\mu \bar{m}_\nu +%
\bar{m}_\mu m_\nu ),  \label{l_covd} \\
n_{\mu ;\nu } &=&-(A^{^{\prime }}/2)n_\mu l_\nu +(A/2r)(m_\mu \bar{m}_\nu +%
\bar{m}_\mu m_\nu ),  \label{n_covd} \\
m_{\mu ;\nu } &=&(A/2r)l_\mu m_\nu -(1/r)n_\mu m_\nu +(\text{cot}\vartheta /%
\sqrt{2}r)(m_\mu m_\nu -m_\mu \bar{m}_\nu ).  \label{m_covd}
\end{eqnarray}
In order to clearly see the two-fluid description we introduce a timelike
unit velocity vector $\hat{v}^\mu $ and three unit spacelike vectors $\hat{r}%
^\mu $, $\hat{\vartheta}^\mu $, $\hat{\varphi}^\mu $ such that 
\end{mathletters}
\[
g_{\mu \nu }=\hat{v}_\mu \hat{v}_\nu -\hat{r}_\mu \hat{r}_\nu -\hat{\vartheta%
}_\mu \hat{\vartheta}_\nu -\hat{\varphi}_\mu \hat{\varphi}_\nu . 
\]
The unit vectors are defined by 
\begin{mathletters}
\begin{eqnarray}
\hat{v}_\mu dx^\mu &=&A^{1/2}du+A^{-1/2}dr,\ \ \ \ \hat{v}^\mu \partial _\mu
=A^{-1/2}\partial _u,  \label{vteta} \\
\hat{r}_\mu dx^\mu &=&A^{-1/2}dr,\ \ \ \ \ \ \ \ \ \ \hat{r}^\mu \partial
_\mu =A^{-1/2}\partial _u-A^{1/2}\partial _r,  \label{vtetb} \\
\hat{\vartheta}_\mu dx^\mu &=&rd\vartheta ,\ \ \ \ \ \ \ \ \ \ \ \ \ \ \hat{%
\vartheta}^\mu \partial _\mu =-r^{-1}\partial _\vartheta ,  \label{vtetc} \\
\hat{\varphi}_\mu dx^\mu &=&r\text{sin}\vartheta d\varphi ,\ \ \ \ \ \ \ \ 
\hat{\varphi}^\mu \partial _\mu =-(r\text{sin}\vartheta )^{-1}\partial
_\varphi .  \label{vtetd}
\end{eqnarray}
$\hat{v}^\mu $ is hypersurface-orthogonal, i.e. $\hat{v}_{[\mu ;\nu }\hat{v}%
_{\alpha ]}=0$, with $h_{\mu \nu }$ the first fundamental form of the
hypersurface. Since $\hat{v}_\mu dx^\mu =f(u,r)dt$, the components of $%
h_{\mu \nu }$ show explicitly that $\hat{v}^\mu $ lies along $t=const$ time
lines: 
\end{mathletters}
\begin{eqnarray}
h_{\mu \nu }dx^\mu dx^\nu &=&(g_{\mu \nu }-\hat{v}_\mu \hat{v}_\nu )dx^\mu
dx^\nu  \label{hmet} \\
&=&-A^{-1}dr^2-r^2(d\vartheta ^2+\text{sin}^2\vartheta d\varphi ^2). 
\nonumber
\end{eqnarray}
The kinematics of the $\hat{v}^\mu $ flow are described by 
\begin{equation}
\hat{v}_{\ ;\nu }^\mu =a^\mu \hat{v}_\nu +\sigma _{\ \nu }^\mu -(\Theta /3)(%
\hat{r}^\mu \hat{r}_\nu +\hat{\vartheta}^\mu \hat{\vartheta}_\nu +\hat{%
\varphi}^\mu \hat{\varphi}_\nu ),  \label{vflow}
\end{equation}
where 
\begin{mathletters}
\begin{eqnarray}
a^\mu &=&[\dot{m}/r+A\partial _r(m/r)]A^{-3/2}\hat{r}^\mu ,  \label{va} \\
\sigma _{\ \nu }^\mu &=&(\Theta /3)(-2\hat{r}^\mu \hat{r}_\nu +\hat{\vartheta%
}^\mu \hat{\vartheta}_\nu +\hat{\varphi}^\mu \hat{\varphi}_\nu ),  \label{vb}
\\
\Theta &=&(\dot{m}/r)A^{-3/2}.  \label{vc}
\end{eqnarray}
The Einstein tensor can now be written as a two-fluid system: 
\end{mathletters}
\begin{eqnarray}
G_{\mu \nu } &=&(2\dot{m}/r^2)l_\mu l_\nu -(2m^{\prime }/r^2)(\hat{v}_\mu 
\hat{v}_\nu -\hat{r}_\mu \hat{r}_\nu )  \label{ein3} \\
&&+(m^{\prime \prime }/r)(\hat{\vartheta}_\mu \hat{\vartheta}_\nu +\hat{%
\varphi}_\mu \hat{\varphi}_\nu ).  \nonumber
\end{eqnarray}
Spherical symmetry allows the function $m(u,r)$ to be identified as the mass
within two-surfaces of constant $u$ and $r$, and invariantly defined from
the sectional curvature \cite{m&s} of those surfaces: 
\begin{equation}
-2m/r^3=R_{\alpha \beta \mu \nu }\hat{\vartheta}^\alpha \hat{\varphi}^\beta 
\hat{\vartheta}^\mu \hat{\varphi}^\nu .  \label{mdef}
\end{equation}

\section{STRING\ FLUID}

The string bivector is defined by 
\[
\Sigma ^{\mu \nu }=\epsilon ^{BC}\frac{\partial x^\mu }{\partial x^B}\frac{%
\partial x^\nu }{\partial x^C},\ \ \ (B,C)=(0,1)\ \text{or}\ (2,3). 
\]
Spherical symmetry demands that the averaged string bivector will have a
world-sheet in either the ($u,r$) or ($\vartheta ,\varphi $) plane. The
condition that the world-sheets are timelike, i.e. $\gamma :=\frac 12\Sigma
^{\mu \nu }\Sigma _{\mu \nu }<0$, implies that only the $\Sigma _{ur}$
component is non-zero. It is useful to write $\Sigma ^{\mu \nu }$ in terms
of unit vectors 
\begin{equation}
\Sigma ^{\mu \nu }=\hat{r}^\mu \hat{v}^\nu -\hat{v}^\mu \hat{r}^\nu .
\label{stringvec}
\end{equation}
It is now clear that $\Sigma ^{\mu \alpha }\Sigma _\alpha ^{\ \nu }=\hat{v}%
^\mu \hat{v}^\nu -\hat{r}^\mu \hat{r}^\nu $. We follow Letelier \cite{let1},%
\cite{let2} and write a string energy-momentum tensor by analogy with one
for a perfect fluid 
\[
T_{\mu \nu }^{fluid}=\rho u_\mu u_\nu -ph_{\mu \nu }, 
\]
where $h_{\ \nu }^\mu =\delta _{\ \nu }^\mu -u^\mu u_\nu $, $\ h_{\ \nu
}^\mu u^\nu =0$. The string energy-momentum is given by 
\begin{equation}
T_{\mu \nu }^{string}=\rho (-\gamma )^{1/2}\hat{\Sigma}_\mu ^{\ \alpha }\hat{%
\Sigma}_{\alpha \nu }-p_{\perp }H_{\mu \nu },  \label{tstring}
\end{equation}
where $H_{\ \nu }^\mu =\delta _{\ \nu }^\mu -\hat{\Sigma}^{\mu \alpha }\hat{%
\Sigma}_{\alpha \nu }$, $\ H_{\ \nu }^\mu \hat{\Sigma}^{\nu \beta }=0$.
Although here $\gamma =-1$, we have kept $\gamma $ explicit in (\ref{tstring}%
) and written $\hat{\Sigma}^{\mu \nu }:=(-\gamma )^{-1/2}\Sigma ^{\mu \nu }$
to show how $\hat{\Sigma}^{\mu \nu }$ is made invariant to
reparameterizations of the world-sheets \cite{let1}.

Einstein's field equations allow the matter portion of $G_{\mu \nu }$ in Eq.(%
\ref{ein3}) to be identified as a string fluid: 
\begin{eqnarray}
T_{\mu \nu } &=&T_{\mu \nu }^{rad}+T_{\mu \nu }^{string},  \label{energymom}
\\
&=&\psi l_\mu l_\nu +\rho \hat{v}_\mu \hat{v}_\nu +p_r\hat{r}_\mu \hat{r}%
_\nu +p_{\perp }(\hat{\vartheta}_\mu \hat{\vartheta}_\nu +\hat{\varphi}_\mu 
\hat{\varphi}_\nu ).  \nonumber
\end{eqnarray}
Thus 
\begin{mathletters}
\begin{eqnarray}
4\pi \psi  &=&-\dot{m}/r^2,  \label{psi} \\
4\pi \rho  &=&-4\pi p_r=m^{\prime }/r^2,  \label{rho} \\
8\pi p_{\perp } &=&-m^{\prime \prime }/r.  \label{pperp}
\end{eqnarray}
Since the contracted Bianchi identities are satisfied for arbitrary $m(u,r)$%
, it follows that the equations of motion $T_{\ ;\nu }^{\mu \nu }=0$ are
identically satisfied for the components of $T_{\mu \nu }$ given in Eq.(\ref
{energymom}).

The components of the contracted Bianchi identities which do not vanish
because of spherical symmetry but rather because of the explicit form $%
A=1-2m(u,r)/r$ are
\end{mathletters}
\[
l_\mu G_{\ ;\nu }^{\mu \nu }=-\nabla _\nu [(2\Phi _{11}+R/4)l^\nu ]-G^{\mu
\nu }l_{\mu ;\nu }
\]
and
\[
n_\mu G_{\ ;\nu }^{\mu \nu }=-\nabla _\nu [(2\Phi _{11}+R/4)n^\nu +\Phi
_{22}l^\nu ]-G^{\mu \nu }n_{\mu ;\nu }\ .
\]

\section{STATIC FLUID MODELS}

Static models are useful in developing insights about time dependent fluids.
We consider two static models that are equivalent to the general
stress-energy tensor described in Eq.(\ref{energymom}) when it is
time-independent. The first is a static isotropic string fluid, and the
second provides a two-fluid interpretation of (\ref{energymom}) and an
interpretation of the transverse stress.

\subsection{Isotropic string fluid}

Consider the stress-energy tensor in Eq.(\ref{energymom}) with static mass
function $m(r)$ and with $p_r=p_{\perp }$: 
\begin{equation}
T_{\mu \nu }^{iso}=-p_r(\hat{v}_\mu \hat{v}_\nu -\hat{r}_\mu \hat{r}_\nu -%
\hat{\vartheta}_\mu \hat{\vartheta}_\nu -\hat{\varphi}_\mu \hat{\varphi}_\nu
).  \label{iso_mom}
\end{equation}
This is clearly an isotropic cloud of strings with equation of state $\rho
+p_r=0$. The mass is determined by the constraint $p_r=p_{\perp }$ or 
\begin{equation}
\frac{m^{\prime \prime }}{2r}=\frac{m^{\prime }}{r^2}.  \label{m_constraint}
\end{equation}
The solution of Eq.(\ref{m_constraint}), in the more general time-dependent
case with null radiation fluid, is 
\begin{equation}
m(u,r)=r^3c_1(u)+c_2(u).  \label{mass_r3}
\end{equation}
For the static case we have $c_1$ and $c_2$ constant. As can be seen from
energy-momentum Eq.(\ref{iso_mom}) the isotropic string cloud is in an
Einstein spacetime.

\subsection{Static two-fluid model}

One can use two different four-velocities, $\hat{u}_\mu $ and $\hat{w}_\mu $%
, to write a two-fluid model: 
\begin{equation}
T_{\mu \nu }^{2fluid}=\rho _2\hat{w}_\mu \hat{w}_\nu +\rho _1\hat{u}_\mu 
\hat{u}_\nu +p_r\hat{r}_\mu \hat{r}_\nu .  \label{enmom_2f}
\end{equation}
The fluid with $\rho _2$ is dust and the other is a fluid with a radial
stress. ($\dot{m}=0$ and there is no Vaidya radiation fluid.) Letelier \cite
{let3} has described a procedure for casting two-fluid stress-energies into
the form of an anisotropic fluid. His method is not adapted for the string
fluid equation of state. As a variation of his method, we transform the
fluid velocities to create two un-normalized vectors: $V_\mu $ timelike and $%
Y_\mu $ spacelike. 
\begin{eqnarray}
\sqrt{\rho _1}\ V_\mu &=&\sqrt{\rho _1}\ \text{cos}\alpha \ \hat{u}_\mu +%
\sqrt{\rho _2}\ \text{sin}\alpha \ \hat{w}_\mu  \label{rotvel_1} \\
\sqrt{\rho _2}\ Y_\mu &=&-\sqrt{\rho _1}\ \text{sin}\alpha \ \hat{u}_\mu +%
\sqrt{\rho _2}\ \text{cos}\alpha \ \hat{w}_\mu .  \label{rotvel_2}
\end{eqnarray}
This transformation obeys 
\begin{equation}
\rho _1\hat{u}_\mu \hat{u}_\nu +\rho _2\hat{w}_\mu \hat{w}_\nu =\rho _1V_\mu
V_\nu +\rho _2Y_\mu Y_\nu .  \label{fluid_map}
\end{equation}
Since $Y_\mu $ is spacelike, Eq.(\ref{rotvel_2}) is valid only for $\hat{u}%
_\mu \neq \hat{w}_\mu $ and $\alpha \neq n(\pi /2)$. The parameter $\alpha $
is defined by $V_\mu Y^\mu =0$: 
\[
\text{cot}2\alpha =\frac{\rho _1-\rho _2}{2\sqrt{\rho _1\rho _2}\ \hat{u}%
_\mu \hat{w}^\mu }. 
\]
The stress-energy $T_{\mu \nu }^{2fluid}$ can be written as 
\begin{equation}
T_{\mu \nu }^{transf}=\rho \hat{V}_\mu \hat{V}_\nu +p_r\hat{r}_\mu \hat{r}%
_\nu +p_{\perp }\hat{Y}_\mu \hat{Y}_\nu  \label{enmom_trsf}
\end{equation}
with 
\begin{equation}
\rho :=V_\alpha V^\alpha \ \rho _1,\ \ \ \ p_{\perp }:=-Y_\alpha Y^\alpha \
\rho _2.  \label{new_prho}
\end{equation}
The transformation in Eq.(\ref{fluid_map}) constrains the new density and
transverse pressure to obey 
\begin{equation}
\rho _1+\rho _2=\rho -p_{\perp }.  \label{map_constraint}
\end{equation}
$T_{\mu \nu }^{transf}$ has the same form as $T_{\mu \nu }^{string}$ in Eq.(%
\ref{energymom}).

If the two fluids have the same velocity, $\hat{u}_\mu =\hat{w}_\mu $, then
the two-fluid stress-energy becomes a single fluid with a radial stress. 
\begin{equation}
T_{\mu \nu }^{2fluid}\rightarrow T_{\mu \nu }^{1fluid}=(\rho _1+\rho _2)\hat{%
u}_\mu \hat{u}_\nu +p_r\hat{r}_\mu \hat{r}_\nu .  \label{enmom_1f}
\end{equation}
The density of this string fluid is just the sum of the two densities. In
the case $\hat{u}_\mu =\hat{w}_\mu $ the transverse pressure is zero. The
transverse pressure reflects the different velocities of the fluids in the
two-fluid model. When the transverse pressure is zero then $m=c_1r+c_2.$

Using a multi-fluid model, one can also describe the radial stress as
reflecting a velocity difference between two dusts so that the complete
stress-energy tensor (\ref{energymom}) could be described by a three-fluid
dust model with all of the dusts moving with different velocities. Which of
the interpretations discussed in this section is most likely depends on the
actual physical situation being modeled.

\section{DIFFUSIVE TRANSPORT}

Diffusion has been a seminal process for the development of our
understanding of many modern systems. Since the work of Einstein \cite
{albert}, Smoluchowski \cite{smol}, and Langevin \cite{langevin}, the ideas
implicit in diffusion have found new application in areas as diverse as the
behavior of stock option values to cosmic strings. Vilenkin \cite{vilen} has
introduced diffusion into the description of cosmic strings by
characterizing string evolution as the formation of Brownian trajectories.
This has also been discussed by Bennett \cite{dpben}. Diffusion is playing
an important part in the growth of our understanding of fluctuations in
quantum gravity \cite{percival1},\cite{percival2} and in the very early life
of our universe. For example, Watabiki \cite{watabiki} has used the
classical diffusion equation to characterize diffusion times in fractal
quantum gravity \cite{jam1},\cite{jam2}. Diffusion may also play a role in
eternal inflation models where inflation field fluctuations can be modeled
as random walks \cite{borde&vilen}. The path integral technique, developed
by Norbert Weiner \cite{wg&jr} to describe diffusive processes, has become
an essential part of the modern view of quantum mechanics.

String collisions, unlike point particle collisions, do not occur at a
single spacetime point (interaction vertex). The two-surface picture of
strings requires the collision (interaction) to be a curve on a world-sheet.
Observers in different Lorentz frames will see the interaction occurring at
different points along the curve. Statistically, the coarse-grained picture
of phase space for a set of collisions hides the lack of Lorentz invariance
of a single collision. We assume string diffusion is like point particle
diffusion in that the number density diffuses from higher numbers to lower
according to 
\begin{equation}
\partial _un={\cal D}\ \nabla ^2n  \label{ndifu}
\end{equation}
where $\nabla ^2=r^{-2}(\partial /\partial r)r^2(\partial /\partial r)$ and $%
{\cal D}$ is the positive coefficient of self-diffusion (which we henceforth
take to be constant). Classical transport theory derives the diffusion
equation by starting with Fick's law 
\begin{equation}
\vec{J}_{(n)}=-{\cal D}\vec{\nabla}n  \label{fick}
\end{equation}
where $\vec{\nabla}$ is a purely spatial gradient. Then 4-current
conservation $J_{(n);\mu }^\mu =0,$ where 
\begin{eqnarray}
J_{(n)}^\mu \partial _\mu &=&(n,\vec{J}_{(n)})  \label{jvec} \\
&=&n\partial _u-{\cal D}(\partial n/\partial r)\partial _r,  \nonumber
\end{eqnarray}
yields the diffusion equation (\ref{ndifu}). We label the 4-current $J_{(n)}$
to indicate $n$ diffusion but we could have also written $J_{(\rho )}$ since
the string number density and string fluid density must be related by $\rho
=M_sn$ where $M_s$ is the constant mass of the string species. $M_s$ must be
a multiple of the Planck mass since it is only over Planck length scales
that point particles resolve into strings.

By rewriting the $T_{\mu \nu }$ components (\ref{psi}) and (\ref{rho}) as $%
\dot{m}=-4\pi r^2\psi $ and $m^{\prime }=4\pi r^2\rho ,$ we can write the
integrability condition for $m$ as 
\begin{equation}
\dot{\rho}+r^{-2}\partial _r(r^2\psi )=0.  \label{rhocontinuity}
\end{equation}
If we compare the diffusion equation (\ref{ndifu}) ($n$ replaced by $\rho $) 
\begin{equation}
\dot{\rho}={\cal D}\ r^{-2}\partial _r(r^2\partial \rho /\partial r)
\label{rhodifu}
\end{equation}
with $\dot{\rho}$ in Eq.(\ref{rhocontinuity}) we obtain 
\begin{equation}
\dot{m}=4\pi {\cal D}\ r^2\partial \rho /\partial r.  \label{mdot2}
\end{equation}
Thus solving the diffusion equation for $\rho $ and then integrating those
solutions to obtain $m$ provides exact Einstein solutions which can be
interpreted as either anisotropic fluids or diffusing string fluids.

There are some analytic solutions of Eq.(\ref{rhodifu}): 
\begin{eqnarray}
\rho &=&\rho _0+k_1/r,  \label{solna} \\
\rho &=&\rho _0+(k_2/6)r^2+k_2{\cal D}u,  \label{solnb} \\
\rho &=&\rho _0+k_3({\cal D}u)^{-3/2}\text{exp}[-r^2/(4{\cal D}u)],
\label{solnc} \\
\rho &=&\rho _0+(k_6/r)\text{exp}(-k_4^2{\cal D}u)[\text{sin}(k_4r)+k_5\text{%
cos}(k_4r)].  \label{solnd}
\end{eqnarray}
Solutions (\ref{solna}) and (\ref{solnc}) appear in \cite{ed&jean}. In
addition to the solutions above, we have obtained a number of analytic
solutions using well known similarity techniques. Those solutions will be
presented elsewhere. The physical behavior of the density solutions provides
a variety of atmospheric models.

Solution (\ref{solna}) describes an atmosphere with a simple drop off in
radius and no time dependence. $k_1$ must be positive to avoid negative
densities. $\rho _0$ is the density at spatial infinity.

The density described by equation (\ref{solnb}) has two very different
behaviors. For $k_2>0$, the density is not physically realistic. It
increases with radius and grows with time. However, for $k_2<0$, the density
decreases with radius and has a zero indicating a bounded string atmosphere.
Since we are working with a string fluid whose equation of state is $\rho
=-p_r$, the boundary also has zero radial pressure. The position of the
boundary moves inward as ${\cal D}u$ increases; the extent of the
atmospheric shell decreases with time.

The third density solution, Eq.(\ref{solnc}), requires $k_3>0$ for positive
densities. The solution begins with a high central density which falls off
with radius. As time progresses the density decreases to the constant value $%
\rho _0$, the density at spatial infinity.

The fourth density solution, Eq.(\ref{solnd}), models a complete array of
atmospheric behaviors. For example, the parameter choices $k_6/\rho _0=10$, $%
k_4=k_5=1$, describe an atmosphere that starts with a string boundary at $%
{\cal D}u=1$ and as time progresses becomes an unbounded string cloud. Other
parameter choices model atmospheric shells which are always unbounded.
Parameters $k_5$ and $k_6$ must be positive for positive densities.

\section{DIFFUSIVE\ MASS SOLUTIONS}

\subsection{Analytic solutions}

Upon integrating $m^{\prime }=4\pi r^2\rho $ and $\dot{m}=4\pi {\cal D}\
r^2\partial \rho /\partial r$ we obtain , in the same order as the densities
above, the following 
\begin{eqnarray}
m(u,r) &=&m_0+(4\pi /3)r^3\rho _0+2\pi k_1(r^2-2{\cal D}u),  \label{mone} \\
m(u,r) &=&m_0+(4\pi /3)r^3\rho _0+(4\pi /3)r^3k_2({\cal D}u+r^2/10),
\label{mtwo} \\
m(u,r) &=&m_0+(4\pi /3)r^3\rho _0+16\pi k_3[-\eta \text{exp}(-\eta ^2)+(%
\sqrt{\pi }/2)\text{erf}(\eta )],  \label{mtre} \\
m(u,r) &=&m_0+(4\pi /3)r^3\rho _0+(4\pi k_6/k_4^2)\text{exp}(-k_4^2{\cal D}%
u)B,  \label{mfour}
\end{eqnarray}
where $\eta :=r\,(4{\cal D}u)^{-1/2}$, and $\ B=\ $sin$(k_4r)-k_4r$cos$%
(k_4r)+k_5[$cos$(k_4r)+k_4r$sin$(k_4r)].$ Solutions (\ref{mone}) and (\ref
{mtre}) appear in \cite{ed&jean}.

If parameters $k_1$, $k_2$, $k_3$, and $k_6$ are zero then all the mass
solutions above become the static solution $m_0+(4\pi /3)r^3\rho _0$. This
is the mass for the isotropic string fluid described by Eq.(\ref{mass_r3}).

\subsection{Interpreting the mass}

Diffusive transport has a conserved four-current for the density 
\[
J^\mu \partial _\mu =\rho \partial _u-{\cal D}(\partial \rho /\partial
r)\partial _r 
\]
which can be written in terms of the null tetrad 
\begin{equation}
J^\mu =\rho n^\mu +[(A/2)\rho -{\cal D}(\partial \rho /\partial r)]l^\mu
\label{jnull}
\end{equation}
When $J_{;\mu }^\mu =0$ is integrated over a four-volume, Stokes theorem
casts the integral onto the bounding three-surfaces: 
\begin{equation}
\int\limits_{R_4}J_{,\mu }^\mu \sqrt{-g}d^4x=\oint\limits_{\partial
R_4}J^\mu \sqrt{-g}dS_\mu  \label{jstokes}
\end{equation}
where we integrate over $u=const$ null three-surfaces ${\cal N}_2$ and $%
{\cal N}_{1\text{ }}$with $dS_\mu =l_\mu drd\vartheta d\varphi $, and $%
r=const$ timelike three-surfaces $\Sigma _2$ and $\Sigma _1$ with $dS_\mu
=(n_\mu -\frac A2l_\mu )dud\vartheta d\varphi $. The four-volume $R_4$ can
be pictured as a tube surrounding a timelike cylinder containing the central
source. An orthogonal cross-section of $R_4$ would have sides ${\cal N}_1$, $%
{\cal N}_2$, $\Sigma _1$, $\Sigma _2$, forming a rhomboid with $\Sigma _1$
bounding $R_4$ away from the central source. Thus 
\[
\int\limits_{{\cal N}_2-{\cal N}_1}J^\mu l_\mu \sqrt{-g}drd\vartheta
d\varphi +\int\limits_{\Sigma _2-\Sigma _1}J^\mu (n_\mu -\frac A2l_\mu )%
\sqrt{-g}dud\vartheta d\varphi =0 
\]
with specific form 
\begin{equation}
\int\limits_{{\cal N}_2-{\cal N}_1}\rho \sqrt{-g}drd\vartheta d\varphi
-\int\limits_{\Sigma _2-\Sigma _1}{\cal D}(\partial \rho /\partial r)\sqrt{-g%
}dud\vartheta d\varphi =0.  \label{j_boundary}
\end{equation}

To first understand the string mass, we examine a static string fluid.
Metric (\ref{met1}) includes static string fluids when $m(u,r)$ is
restricted to $m(r)$. We can use the density solution Eq.(\ref{solna}) with $%
{\cal D}=0$ so that there is no diffusion. For $\rho =\rho _0+k_1/r$, we
have 
\begin{equation}
4\pi \int\limits_{{\cal N}_2-{\cal N}_1}(\rho _0r^2+k_1r)dr=0.
\label{static_rho}
\end{equation}
Thus the string mass is static with $m_{string}(r)=\int\limits_{{\cal N}%
_1}=\int\limits_{{\cal N}_2}:$%
\begin{equation}
m_{string}(r)=m_0+4\pi (\frac 13\rho _0r^3+\frac 12k_1r^2)  \label{static_m}
\end{equation}
on null surface ${\cal N}_1$ and at a later time on ${\cal N}_2$. This is
the mass in Eq.(\ref{mone}) with ${\cal D}=0$.

Now consider the case with time dependence where the mass is $m(u,r)$ and
there is a Vaidya fluid of short photons. Again using $\rho =\rho _0+k_1/r$,
we find 
\begin{equation}
4\pi \int\limits_{{\cal N}_2-{\cal N}_1}(\rho _0r^2+k_1r)dr+4\pi
\int\limits_{\Sigma _2-\Sigma _1}k_1{\cal D}du=0  \label{sum_0}
\end{equation}
This yields the total mass of strings and null fluid in Eq.(\ref{mone})
where 
\begin{equation}
m(u,r)=\int\limits_{{\cal N}_2}-\int\limits_{\Sigma _1}=\int\limits_{{\cal N}%
_1}-\int\limits_{\Sigma _2}  \label{total_m}
\end{equation}
We use our knowledge of the static case to identify the string mass as that
part of the total mass integrated over the null three-surface ${\cal N},$
and identify the flux through the timelike surface $\Sigma $ as resulting
from both the energy carried by the Vaidya photons and the diffusing
strings. Photons enter $R_4$ from the central source through $\Sigma _1$
while string bits diffuse through $\Sigma _1$ toward the source, with the
opposite happening at $\Sigma _2$. There the photons leave $R_4$ while
string bits enter. Thus $m(u,r)=m_{string}+m_{flux}$ where 
\begin{eqnarray*}
m_{string} &=&4\pi (\frac 13\rho _0r^3+\frac 12k_1r^2) \\
m_{flux} &=&-4\pi k_1{\cal D}u
\end{eqnarray*}
Of course $m_{string}$ need not be static. The density examples with $\rho
(u,r)$ will have $m_{string}(u,r)$.

\section{HORIZONS}

The topological two-spheres ($\vartheta ,\varphi $) nested in an $r=const$
surface at time $u$ have outgoing null geodesic normal $l^\mu $ and incoming
null geodesic normal $n^\mu $. The spheres become trapped surfaces when both 
$l^\mu r,_\mu $ and $n^\mu r,_\mu $ are positive (negative for +2
signature). The marginally trapped surface is the outer boundary of all
trapped surfaces at time $u$, and the apparent horizon is the time history
of the marginally trapped surface.

Here $l^\mu r,_\mu =1$. The fluid atmosphere has $n^\mu r,_\mu =-A/2$,
negative until $m(u,r)\geq r/2$. At that time the mass has become compact
enough to trap light.

We will analyze mass expression (\ref{mone}) for trapped surfaces. With $%
L_0:=2m_0$, $L_1^{-3}:=(4\pi \rho _0)/(3m_0)$, $L_2^{-2}:=2\pi k_1/m_0$, and 
$1/T_0:=4\pi k_1{\cal D}/m_0$, we search for zeros of the polynomial 
\begin{equation}
\frac{r^3}{L_1^3}+\frac{r^2}{L_2^2}-\frac r{L_0}+1-\frac u{T_0}=0
\label{r_cubic}
\end{equation}
at time $u=const.$ The four parameters ($T_0,L_0,L_1,L_2$) are positive.
Suppose there is a root at $r=\alpha L_0$, $\alpha >0$. Then, at $u=0$, Eq.(%
\ref{r_cubic}) becomes 
\[
\alpha ^3(L_0/L_1)^3+\alpha ^2(L_0/L_2)^2+1-\alpha =0 
\]
or 
\begin{equation}
\alpha ^2(2m_0)^2[\alpha (8\pi \rho _0/3)+2\pi k_1/m_0]+1-\alpha =0.
\label{alpha_eqn}
\end{equation}
If $k_1$ and $\rho _0$ are zero then $\alpha =1$ and we find the
Schwarzschild horizon at $r=2m_0$. For non-zero parameters Eq.(\ref
{alpha_eqn}) has no real roots, which is consistent with the outgoing
short-wavelength photons.

At time $u=0$ and for some short time after $1-u/T_0$ is positive. After
more time $\Delta u,$ $1-u/T_0$ becomes negative and then there is only one
sign change in Eq.(\ref{r_cubic}). Descartes' rule of signs tells us that,
for $u>\Delta u$, there is at most one real root of Eq.(\ref{r_cubic}) and
the fluid will have a trapped surface at $r=2m(u,r)$.

\section{CONCLUSION}

That $\hat{v}^\mu $ is hypersurface orthogonal implies $\hat{v}_\mu dx^\mu
=f(u,r)dt$. If metric (\ref{met1}) were the Schwarzschild metric then $%
A=1-2m_0/r$, $u=t-r-2m_0$ln$(r-2m_0)$, and $\hat{v}_\mu dx^\mu =A^{-1/2}dt$.
Next in simplicity is the Vaidya metric, where $A$ of metric (\ref{met1})
would be $A=1-2m(u)/r$. The Vaidya metric is not static and one cannot
coordinate transform back to Schwarzschild. With the Vaidya metric $r-2m(u)$
is a spacelike hypersurface lying outside the local null cone \cite{ls&m},
unlike the Schwarzschild hypersurface $r-2m_0$ which is null.

The traditional view of the Vaidya metric places it outside a spherical star
which is losing mass via Vaidya's short-wavelength photons. Vacuum
Schwarzschild geometry is joined smoothly to Vaidya at its radiative
boundary.

The system described here with $m(u,r)$ continues to have $\hat{v}^\mu $
hypersurface orthogonal, so the $t$ in $\hat{v}_\mu dx^\mu =f(u,r)dt$ labels
spacelike hypersurfaces with energy-momentum given by Eq.(\ref{energymom})
above. Because of the implicit equation of state, $\rho +p_r=0$, we
interpreted the fluid as an ''atmosphere'' of open strings which form a
string fluid. This view has also been discussed by 't Hooft \cite{thooft}
and Maldacena \cite{mald}. This fluid was taken as a source of
energy-momentum. As a modeling choice, we took the strings to interact
diffusively with energy carried away by short-wavelength photons. It was
necessary to use the approximation of short-wavelength photons since $(2\dot{%
m}/r^2)l_\mu l_\nu $ is not an exact solution of Maxwell's equations.

Quantum field theory requires the quantum vacuum to be Lorentz invariant 
\cite{lima}, which constrains the energy-momentum tensor on a Minkowski
background to have the form $T_{\mu \nu }=\rho \eta _{\mu \nu }$, and the
energy-density to transform as 
\[
\rho ^{\prime }=\frac{\rho +p(v^2/c^2)}{1-(v^2/c^2)} 
\]
under change of inertial frame. The string equation of state $p_r=-\rho $
satisfies the required transformation property and so the vacuum outside a
relativistic star could indeed include a string atmosphere.

This work necessarily forms an incomplete picture of the evolution of the
astrophysical system we model here. Strings may exist at the Planck length
scale, with a large number of them possibly providing a macroscopic
classical string, and forming a visible atmosphere until they lie within a
trapped surface.\\

{\bf ACKNOWLEDGMENTS}

E.N. Glass was partially supported by an NSERC of Canada grant. Computations
were verified using MapleV.4 (Waterloo Maple Software, Waterloo, Ontario)
and GRTensorII rel 1.59 (P. Musgrave, D.Pollney, and K. Lake, Queens
University, Kingston, Ontario).

\end{document}